\documentclass[lettersize,journal]{IEEEtran}
\usepackage{amsmath,amsfonts,amssymb}
\usepackage{array}
\usepackage[caption=false,font=normalsize,labelfont=sf,textfont=sf]{subfig}
\usepackage{textcomp}
\usepackage{stfloats}
\usepackage{url}
\usepackage{verbatim}
\usepackage{graphicx}
\usepackage{hyperref}
\usepackage{bm}
\usepackage{algorithm}
\usepackage{algorithmic}
\usepackage{graphicx}
\def\BibTeX{{\rm B\kern-.05em{\sc i\kern-.025em b}\kern-.08em
    T\kern-.1667em\lower.7ex\hbox{E}\kern-.125emX}}

\usepackage{balance}
\usepackage{booktabs}

\newtheorem{example}{Example}

\begin{document}
\title{Extended Isolation Forest with feature sensitivities}
\author{Illia Donhauzer\thanks{Institute of Mathematics for Industry, Kyushu University, Japan; La Trobe University, Australia; email: \href{i.donhauzer@latrobe.edu.au}{i.donhauzer@latrobe.edu.au}}}


\maketitle

\begin{abstract}
Compared to theoretical frameworks that assume equal sensitivity to deviations in all features of data, the theory of anomaly detection allowing for variable sensitivity across features is less developed. To the best of our knowledge, this issue has not yet been addressed in the context of isolation-based methods, and this paper represents the first attempt to do so. This paper introduces an Extended Isolation Forest with feature sensitivities, which we refer to as the Anisotropic Isolation Forest (AIF). In contrast to the standard EIF, the AIF enables anomaly detection with controllable sensitivity to deviations in different features or directions in the feature space. The paper also introduces novel measures of directional sensitivity, which allow quantification of AIF’s sensitivity in different directions in the feature space. These measures enable adjustment of the AIF’s sensitivity to task-specific requirements. We demonstrate the performance of the algorithm by applying it to synthetic and real-world datasets. The results show that the AIF enables anomaly detection that focuses on directions in the feature space where deviations from typical behavior are more important.


\end{abstract}



\section{Introduction}

The necessity of efficient anomaly detection methods has motivated active research in this area and has, over the last decades, resulted in the development of a wide variety of anomaly detection techniques based on different principles. Statistical approaches assume an underlying probabilistic model of the data and identify anomalies as low-likelihood events \cite{Barnett}. Proximity- and density-based methods, such as $k$-nearest neighbours and the Local Outlier Factor (LOF), detect anomalies based on distances between data points or deviations in local density \cite{Breunig}. Clustering-based techniques regard data points that fit poorly into clusters or belong to small clusters as anomalous \cite{Eskin}.

In recent years, the Isolation Forest \cite{Liu} has
emerged as arguably the most popular anomaly detector
due to its general effectiveness across different benchmarks
and strong scalability \cite{Xu}. Instead of building an explicit model of nominal data, isolation-based anomaly detectors exploit the idea that anomalies are “few and different” and can therefore be easily isolated from nominal (non-anomalous) data points through random partitioning of the feature space.

When performing anomaly detection, deviations in some features of data may be more important than deviations in others. Unfortunately, neither the Isolation Forest nor the further-developed Extended Isolation Forest (EIF) \cite{Hariri} allows for the explicit specification of feature sensitivities (how sensitive the method should be to deviations in different features). However, in this paper, we show that feature sensitivities can be incorporated into the Extended Isolation Forest without the need to devise a fundamentally different method. We achieve this by adjusting the way the Extended Isolation Forest slices the data and isolates anomalous data points.

The main idea behind the proposed approach is that data points with atypical values in certain features can be quicker isolated by appropriately oriented hyperplanes. In the original Extended Isolation Forest, the data are sliced using random hyperplanes whose normal vectors are sampled from a standard multivariate Gaussian distribution that is symmetric in all directions (isotropic). As a result, the method treats data points equally in all directions of the feature space. In contrast, the proposed Anisotropic Isolation Forest (AIF) slices the data using hyperplanes with random normal vectors sampled from direction-dependent (anisotropic) distributions. These hyperplanes isolate data points with atypical values in selected features, or data points located in selected directions of the feature space, more quickly, which in turn leads to higher anomaly scores for those data points.

AIF does not simply cancel out features with lower sensitivities by assigning anomaly scores based only on the strongest ones, but considers all the features jointly and assigns anomaly scores based on their combined sensitivities. Consequently, the AIF exhibits direction-dependent sensitivity, assigning higher anomaly scores to data points in selected directions of the feature space and lower scores in others. One of the key contributions of this paper is the introduction of AIF's measures of directional sensitivity, which quantify how sensitive the AIF is to deviations in different directions of the feature space. This framework enables the adjustment of the AIF to task-specific requirements, allowing it to be made more or less sensitive in selected directions.

This paper is organized as follows. Section~\ref{sec2} reviews the existing methodology, presents motivating examples, introduces the proposed AIF algorithm, and defines the AIF’s directional sensitivity measures. Section~\ref{sec:num} applies the AIF to synthetic and real-world datasets, compares its performance with the EIF, analyzes anomaly score heat maps, examines the relationship between measures of directional sensitivity and the assigned anomaly scores, and studies the distributions of detected anomalies under different feature sensitivity scenarios.

\section{Anisotropic isolation forest}
\label{sec2}
\subsection{Existing methodology and the proposed approach}
\label{sec2:sub1}
Both the Isolation Forest (IF) and the Extended Isolation Forest (EIF) consist of two main stages: (i) training, which involves building binary search trees using subsamples of the dataset, and (ii) scoring, which involves traversing these trees for all data points to compute anomaly scores \cite{kmeans}. The EIF generalizes the standard Isolation Forest by allowing more flexible partitioning during tree construction using random hyperplane splits with arbitrary orientations, while in the IF random hyperplanes are axis-aligned.

During the construction of a search tree in the EIF, a random subsample of the data is first selected. At each branching step, a random normal vector $\bm{n} = (n_1, n_2, \ldots, n_d)$
is sampled from the standard multivariate Gaussian distribution $N(\bm{0}, \bm{I})$, where $d$ is the dimensionality of the data (number of features), $\bm{0} = (0, 0, \ldots, 0)$ is the zero vector, and $\bm{I}$ is the $d \times d$ identity matrix. The splitting rule for a data point $\bm{x}$ is then given by
\begin{equation}\label{eq:eif1}
(\bm{x} - \bm{p})' \bm{n} < 0,
\end{equation}
where $\bm{p}$ is a random intercept chosen uniformly from the range of the data points considered at the current branching step. If the inequality~\eqref{eq:eif1} holds, the point is sent to the left branch; otherwise, it is directed to the right. This recursive partitioning continues until each data point in the subsample is isolated in its own node or a prespecified maximum depth of the search tree is reached. An ensemble of such search trees constitutes the EIF.

During the scoring phase, a candidate observation is passed down each search tree in the forest. Anomalous points are typically isolated at shallower depths than nominal points. The anomaly score of a data point $\bm{x}$ is computed as
\begin{equation}
s(\bm{x}) = 2^{-\frac{E(h(\bm{x}))}{c(n)}},
\end{equation}
where $E(h(\bm{x}))$ denotes the average path length of $\bm{x}$ across all trees, and $c(n)$ is a normalization constant defined as the expected path length of an unsuccessful search in a binary search tree. It is computed as

\begin{equation}
\label{const}
c(n) = 2H(n-1) - \frac{2(n-1)}{n},
\end{equation}
such that $H(n) = \ln(n) + 0.5772156649$ (the latter number is Euler-Mascheroni constant), and $n$ is the number of points used in the construction of the trees.

Because the EIF uses random hyperplane splits rather than axis-aligned cuts, it is more efficient that the IF at separating points near oblique or complex boundaries \cite{Hariri}. It makes the EIF particularly effective for high-dimensional data with complex dependencies between features, where axis-parallel partitioning would require many splits to isolate data points.

In contrast to the standard EIF, we consider the EIF constructed using hyperplanes whose normal vectors are sampled from anisotropic multivariate Gaussian distributions \( N(\bm{0}, \bm{A}) \), where \( \bm{A} \) is not necessarily the identity matrix, or mixtures of Gaussian distributions.

\begin{figure}[htb!]
\centering
\subfloat[]{\includegraphics[width=1.1in]{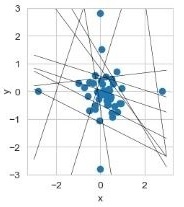}%
\label{fig1.1}}
\hfil
\subfloat[]{\includegraphics[width=1.1in]{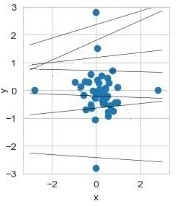}%
\label{fig1.2}}
\hfil
\subfloat[]{\includegraphics[width=1.1in]{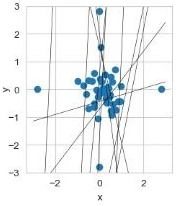}%
\label{fig1.3}}
\caption{Partitions of data.}
\label{fig1}
\end{figure}

\begin{figure}[htb!]
\centering
\subfloat[]{\includegraphics[width=1.1in]{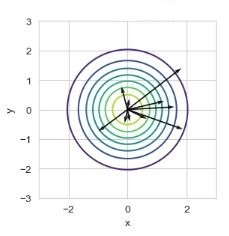}%
\label{fig2.1}}
\hfil
\subfloat[]{\includegraphics[width=1.1in]{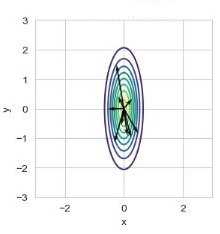}%
\label{fig2.2}}
\hfil
\subfloat[]{\includegraphics[width=1.1in]{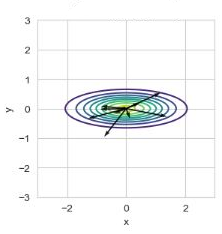}%
\label{fig2.3}}
\caption{Distributions and sampled normal vectors.}
\label{fig2}
\end{figure}

To illustrate the motivation for this idea, consider Fig.~\ref{fig1}, which shows three random line partitions of a simple two-dimensional dataset. Fig.~\ref{fig2} presents the theoretical distributions of the normal vectors of the random lines in the corresponding partitions shown in Fig.~\ref{fig1}, together with samples drawn from these distributions. 

In the first partition, normal vectors are sampled from \( N(\bm{0}, \bm{I}) \), so that the angles between the lines and the \( x \)-axis are uniformly distributed over the interval \( [0, \pi] \). In the second partition, the random lines are more frequently aligned with the \( x \)-axis. One can observe that in the second partition, the random lines fail to isolate data points with extreme \( x \)-values from the rest of the data but can effectively isolate points with extreme \( y \)-values. In contrast, in the first partition, data points with extreme \( x \)-values or extreme \( y \)-values are isolated by approximately equal numbers of random lines. The third partition reverses this anisotropy: the random lines are aligned more frequently with the \( y \)-axis. Data points with extreme \( x \)-values are isolated by one or a few lines, whereas data points with extreme \( y \)-values are either not isolated or are isolated only after many splits.

The above examples show that, in partitions constructed using random lines that are more frequently oriented in certain directions, some data points are separated more quickly (by one or a few random lines), while others are separated more slowly, requiring a larger number of lines in the partition. More precisely, when the random lines are more aligned (i.e., more parallel) with a given axis, data points with atypical values along that axis tend to require more random lines to be isolated. In contrast, when the random lines are less aligned (i.e., more perpendicular) with that axis, such data points tend to be isolated more quickly.

Consider again the dataset shown in Fig.~\ref{fig1} and an EIF whose branch cuts are induced by random lines with normal vectors sampled from the distribution shown in Fig.~\ref{fig2.2}. In partitions constructed using such lines, data points with atypical \( y \)-values tend to be separated more quickly, whereas data points with atypical \( x \)-values tend to be separated more slowly. Consequently, search trees constructed  using such lines tend to isolate points with atypical \( y \)-values at shallower depths and points with atypical \( x \)-values at greater depths. As a result, the EIF composed of these search trees becomes more sensitive to data points with atypical \( y \)-values, assigning higher anomaly scores to such points, while being less sensitive to data points with atypical \( x \)-values, which receive comparatively lower anomaly scores. Analogously, an EIF whose branch cuts are induced by random lines with normal vectors sampled from the distribution shown in Fig.~\ref{fig2.3} assigns higher anomaly scores to data points with atypical \( x \)-values and comparatively lower anomaly scores to data points with atypical \( y \)-values.

In what follows, let us refer to the EIF with branch cuts induced by random hyperplanes whose normal vectors are sampled from an anisotropic distribution as the AIF (Anisotropic Isolation Forest) and denote it as AIF$(\bm{A})$ if the normal vectors are sampled from $N(\bm{0}, \bm{A})$.

\begin{figure}[htb!]
\centering
\subfloat[]{\includegraphics[width=1.1in]{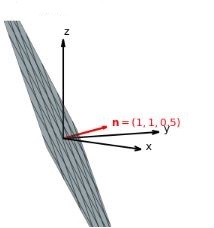}%
\label{fig3.1}}
\hfil
\subfloat[]{\includegraphics[width=1.1in]{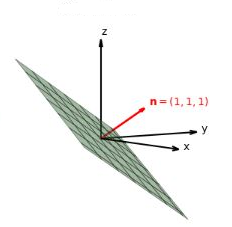}%
\label{fig3.2}}
\hfil
\subfloat[]{\includegraphics[width=1.1in]{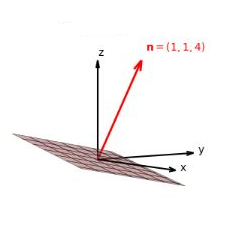}%
\label{fig3.3}}
\caption{Planes and their normal vectors.}
\label{fig3}
\end{figure}

Let \( \bm{A} = \mathrm{diag}(a_1, a_2, \ldots, a_d) \) be a diagonal matrix with entries \( a_1, a_2, \ldots, a_d \). Increasing one of the entries \( a_i \) causes the normal vectors sampled from \( N(\bm{0}, \bm{A}) \) to be more strongly aligned with the corresponding coordinate axis. Consequently, the random hyperplanes defined by these normal vectors become less aligned with that axis. For illustration, see Fig.~\ref{fig3}, where three planes and their corresponding normal vectors are shown. One can observe that, as the \( z \)-component of the normal vector increases, the associated plane becomes increasingly misaligned with the \( z \)-axis.

By adjusting the entries of the diagonal matrix \( \bm{A} = \mathrm{diag}(a_1, a_2, \ldots, a_d) \), one controls the alignment of the hyperplanes with the coordinate axes and, consequently, the sensitivity of AIF(\( \bm{A} \)) to atypical values in different data features. Specifically, AIF(\( \bm{A} \)) becomes more sensitive to atypical values along axes corresponding to larger \( a_i \) and less sensitive to those along axes corresponding to smaller \( a_i \).



When \( \bm{A} \) is not necessarily diagonal, normal vectors sampled from \( N(\bm{0}, \bm{A}) \) are more likely to align with the eigenvectors of \( \bm{A} \) associated with larger eigenvalues. Consequently, AIF(\( \bm{A} \)) assigns higher anomaly scores to data points with atypical values along directions corresponding to eigenvectors with large eigenvalues, and lower anomaly scores to points aligned with eigenvectors associated with smaller eigenvalues. By modifying the eigenvalue \( \lambda_i \) corresponding to an eigenvector \( \bm{g}_i \) of \( \bm{A} \), one adjusts the sensitivity of AIF(\( \bm{A} \)) to atypical values along the direction of \( \bm{g}_i \).

Moreover, \( \bm{A} \) may have its first \( k \) eigenvalues equal, \( \lambda_1 = \lambda_2 = \cdots = \lambda_k \), in this case the distribution \( N(\bm{0}, \bm{A}) \) exhibits increased variance in all directions within the eigenspace spanned by the eigenvectors \( \bm{g}_1, \bm{g}_2, \ldots, \bm{g}_k \). As a result, AIF(\( \bm{A} \)) assigns higher anomaly scores to data points with atypical values along any direction in this eigenspace. Analogously, if the last \( k \) eigenvalues of \( \bm{A} \) are equal, AIF(\( \bm{A} \)) becomes less sensitive along all directions within the corresponding eigenspace.

Thus, AIF(\( \bm{A} \)) can be more or less sensitive to data points with atypical values along different directions in the feature space. In particular, the higher the variance of the distribution \( N(\bm{0}, \bm{A}) \) in a given direction, the higher the anomaly scores that AIF(\( \bm{A} \)) assigns to data points with atypical values along that direction.

\subsection{The case of mixtures of Gaussian distributions}\label{sec:mix}
The sensitivity of AIF($\mathbf{A}$) across a feature space is determined by the anisotropy properties of the distribution $N(\mathbf{0}, \mathbf{A})$. However, this distribution can exhibit only an elliptical form of anisotropy. As a result, the sensitivity patterns of AIF($\mathbf{A}$) derived from $N(\mathbf{0}, \mathbf{A})$ may be insufficient for practical applications, in which more complex sensitivity patterns across the feature space may be required.

Mixtures of Gaussian distributions \cite{Rasmussen}
\[
\sum_{i=1}^{p} \pi_i\, N(\mathbf{0}, \mathbf{A}_i), \quad i = 1,\ldots,p,\; p\in\mathbb{N},
\]
can exhibit complex anisotropy properties that are not necessarily elliptical.
In such mixtures, the coefficients $\pi_i$ represent the probabilities that a vector is sampled from the corresponding Gaussian distribution $N(\mathbf{0}, \mathbf{A}_i)$.
By adjusting the probabilities $\pi_i$ and the covariance matrices $\mathbf{A}_i$, one can construct mixture distributions with highly flexible anisotropy properties.
Consequently, an AIF whose branch cuts are induced by hyperplanes with normal vectors sampled from mixtures of Gaussian distributions can exhibit more complex sensitivity patterns across the feature space—patterns that cannot be achieved in the purely Gaussian case.

\subsection{Measures of directional sensitivity}
\label{sec:sens}
We introduce measures of directional sensitivity that allows to quantify the sensitivity of AIF(\( \bm{A} \)) in different directions in a feature space. In what follows, we assume that the covariance matrix \( \bm{A} \) has unit spectral radius (i.e., its largest eigenvalue equals \( 1 \)). This requirement does not affect the performance of the AIF, but is introduced for normalization purposes, which will be seen later.

As discussed in section~\ref{sec2:sub1}, the higher the variance of the distribution \( N(\bm{0}, \bm{A}) \) in a given direction, the more sensitive AIF(\( \bm{A} \)) becomes to atypical values along that direction. Thus, we quantify the sensitivity of AIF(\( \bm{A} \)) in a given direction by measuring the spread of the distribution \( N(\bm{0}, \bm{A}) \) along that direction.

For the multivariate Gaussian distribution \( N(\bm{0}, \bm{A}) \), the contour surfaces (or contour lines in \( \mathbb{R}^2 \)) are defined by the equations (since the exponential function is monotonic)
\begin{equation}\label{eq:ell}
\bm{x}' \bm{A}^{-1} \bm{x} = c, \quad c > 0.
\end{equation}
The ellipsoids defined by \eqref{eq:ell} are more elongated in directions corresponding to higher variance of the distribution \( N(\bm{0}, \bm{A}) \). Thus, measuring the elongation of these ellipsoids in a given direction provides a measure of the spread of \( N(\bm{0}, \bm{A}) \) in that direction. Since these ellipsoids are homothetic for all \( c > 0 \), it is sufficient to consider the elongation of a single ellipsoid, for example the one corresponding to \( c = 1 \).

The elongation of an ellipsoid along a given axis can be quantified by the length of the ellipsoid's projection onto that axis. In particular, the half-length of the projection of the ellipsoid defined by \eqref{eq:ell} (with \( c = 1 \)) onto the axis directed by a vector \( \bm{n} \in \mathbb{S}^{d-1} \), where \(\mathbb{S}^{d-1} := \{ \bm{n} \in \mathbb{R}^d : \| \bm{n} \| = 1 \}
\) denotes the unit sphere in \( \mathbb{R}^d \), admits the simple expression \( \sqrt{\bm{n}' \bm{A} \bm{n}} \). For completeness of the exposition, the proof is provided in the Appendix \ref{sect:append}.

We therefore define the sensitivity of AIF(\( \bm{A} \)) in the direction \( \bm{n} \in \mathbb{S}^{d-1} \) as
\begin{equation}
\label{eq:sens1}
\alpha(\bm{n}) := \sqrt{\bm{n}' \bm{A} \bm{n}}.
\end{equation}

\begin{example}
Let \( \bm{I} \) denote the identity matrix. Then AIF(\( \bm{I} \)) reduces to the standard EIF and exhibits constant unit sensitivity in all directions, that is,
\[
\alpha(\bm{n}) = 1 \quad \text{for all } \bm{n} \in \mathbb{S}^{d-1}.
\]

\end{example}

\begin{example}
Let \( \bm{A} = \mathrm{diag}(a_1, a_2, \ldots, a_d) \). Then
\[
\alpha(\bm{e}_i) = \sqrt{a_i},
\]
where \( \bm{e}_i \) denotes the unit vector along the \( i \)-th coordinate axis. Thus, \( \sqrt{a_i} \) can be interpreted as the sensitivity of the AIF($\bm{A}$) to deviations in the corresponding feature.

\end{example}

\begin{example}
Let \( \bm{g}_1, \bm{g}_2, \ldots, \bm{g}_d \) be the eigenvectors of \( \bm{A} \) (with unit norm), with corresponding eigenvalues \( \lambda_1, \lambda_2, \ldots, \lambda_d \). Then
\[
\alpha(\bm{g}_i) = \sqrt{\lambda_i}, \quad i = 1, 2, \ldots, d.
\]

\end{example} 

From \eqref{eq:sens1} and Example above, one can see the motivation for requiring the covariance matrix \( \bm{A} \) to have unit spectral radius. Indeed, it holds that
\begin{equation}\label{eq:normalization}
\max_{\bm{n} \in \mathbb{S}^{d-1}} \alpha(\bm{n}) = \max_i \sqrt{\lambda_i} = 1.
\end{equation}
Thus, this requirement normalizes \( \alpha(\cdot) \) to take values in \( [0,1] \).

As an additional tool, we introduce another sensitivity measure that quantifies the sensitivity of AIF(\( \bm{A} \)) over a region of directions \( B \subseteq \mathbb{S}^{d-1} \):
\begin{equation}
\label{tau}
\tau(B) := \frac{1}{|B|} \int_{B} \sqrt{\bm{n}' \bm{A} \bm{n}} \, d\bm{n}.
\end{equation}
It represents the average directional sensitivity \( \alpha(\cdot) \) over the set of directions in \( B \). It is useful in situations where the sensitivity of the AIF needs to be adjusted—either increased or reduced—over an entire region of directions (this measure is increasingly valuable in the case of mixtures of Gaussian distributions, which will be considered below).

The integral in \eqref{tau} can be approximated using Monte Carlo or other numerical approximation methods implemented in systems such as Maple or Wolfram~Alpha.

In the case of normal vectors sampled from mixtures of Gaussian distributions $\sum_{i=1}^p \pi_i N(\bm{0}, \bm{A}_i)$, analogous to \eqref{eq:sens1}, we define the AIF’s measure of directional sensitivity in the direction $\bm{n} \in \mathbb{S}^{d-1}$ as

\begin{equation}\label{eq:sens11} \widetilde\alpha(\bm{n}) :=  \sum\limits_{i=1}^k \pi_i \sqrt{\bm{n}' \bm{A}_i \bm{n}},\end{equation} and, analogously to \eqref{tau}

\begin{equation}\label{tau2} \widetilde\tau(B):=\frac{1}{|B|}\sum\limits_{i=1}^p\pi_i\int\limits_{B} \sqrt{\bm{n}' \bm{A}_i \bm{n}}d\bm{n},\end{equation} for $B\subseteq\mathbb{S}^{d-1}.$

Note that $\widetilde\alpha(\cdot)$ does not necessarily satisfy the normalization condition
\begin{equation}
\label{eq:sens_adj}
\max_{\bm{n}\in\mathbb{S}^{d-1}} \widetilde\alpha(\bm{n}) = 1.
\end{equation}

However, the AIF's performance does not change if one samples normal vectors from
\begin{equation}\label{eq:mixc}
\sum_{i=1}^p \pi_i\, N(\bm{0}, c\bm{A}_i)
\end{equation}
for any $c>0$.
Therefore, for normalization purposes, one can sample normal vectors from \eqref{eq:mixc} with a suitably chosen $c>0$ such that \eqref{eq:sens_adj} is satisfied.

\subsection{High dimensional data and extension levels}

In the original EIF, the extension levels were introduced to handle splits in high-dimensional data and to reduce computational overhead. For a dataset with $d$ features, there are $d$ possible extension levels. In the fully extended case ($d-1$ extensions), the hyperplanes may take arbitrary orientations in the $d$-dimensional feature space and can intersect all $d$ axes. If the extension level is set to $k-1$, the hyperplanes are restricted to intersect only $k$ axes. In the original Isolation Forest, the hyperplanes intersect only a single axis, corresponding to $0$ extension levels.

Let us study the effect of extension levels in the AIF. First, consider a random vector
\(
\bm{\omega} = (\omega_1, \omega_2, \ldots, \omega_d) \sim N(\bm{0}, \bm{I}).
\)
If $\theta_i$ denotes the angle between $\bm{\omega}$ and the $i$-th coordinate axis, then
\begin{equation}\label{cos}
\cos(\theta_i) = \frac{\omega_i}{\sqrt{\omega_1^2 + \omega_2^2 + \cdots + \omega_d^2}} .
\end{equation}
When the dimensionality $d$ is large, the denominator in \eqref{cos} tend to take large values, causing $\cos(\theta_i)$, $i=1,\ldots,d$, to concentrate near zero. Consequently, the random vector $\bm{\omega}$ becomes nearly orthogonal to each coordinate axis. This implies that a random hyperplane with normal vector $\bm{\omega}\sim N(\bm{0}, \bm{I})$ becomes increasingly aligned with the coordinate axes as the dimensionality increases. It shows that the behavior of random hyperplanes in high-dimensional spaces differs from that observed in lower-dimensional settings.

Consider a setting in which the data have $d$ features, with $d$ being large, and in which deviations in the first feature are more important than deviations in the other features. One can construct AIF($\bm{A}$), where $\bm{A} = \mathrm{diag}(a_{1}, a_{2}, \ldots, a_{d})$ and $a_{1}$ is chosen to be larger than the other $a_i$, so that the AIF becomes more sensitive to atypical values in the first feature. However, the effect of adjusting $a_{1}$ may be negligible. Indeed, since the denominator in \eqref{cos} tends to take large values, all $\cos(\theta_i)$, $i=1,\ldots,d$, concentrate near zero. Thus, increasing $a_{1}$ does not lead to a significant change in the orientations of the random hyperplanes. Consequently, adjusting $a_1$ may not result in meaningful changes in the anomaly scores assigned by AIF($\bm{A}$). On the other hand, one could choose $a_1$ to be extremely large, so that $\cos(\theta_1)$ tends to take larger values. In this case, however, the other $\cos(\theta_i), \ i=\overline{2,d},$ become even closer to zero, causing the AIF to effectively ignore the corresponding features.


To overcome the problem of a large denominator in \eqref{cos}, one can use the concept of extension levels proposed in the original EIF paper \cite{Hariri}. In the EIF, $k$ levels of extension are achieved by setting, at each branch cut, $d - k - 1$ randomly selected components $\omega_i$ of $\bm{\omega}$ to zero. Thus, the number of nonzero $\omega_i$ terms in the denominator of \eqref{cos} is reduced to $k$, and therefore the denominator takes smaller values. Hyperplanes with these new normal vectors split the data based on the values of the selected $k$ features, taking their relative sensitivities into account more efficiently.

Geometrically, this can be interpreted as projecting the data at each branch cut onto subspaces spanned by randomly selected $k$ axes, and then splitting the projected data using hyperplanes whose normal vectors are projections of normal vectors sampled from $N(\bm{0}, \bm{A})$ onto those subspaces. The resulting normal vectors are not necessarily isotropically oriented, but retain the orientations of the original normal vectors within the projection subspaces. Consequently, hyperplanes with these new normal vectors can split the projected data in $k$-dimensional subspaces taking into consideration the sensitivities of the corresponding features.

Thus, setting lower extension levels allows to avoid the  large denominator problem in \eqref{cos} by reducing the dimensionality of the data at each branch cut, projecting it onto random subspaces and splitting the projected data by hyperplanes that retain the orientations of the original hyperplanes in the projection subspaces.

\subsection{The Algorithm}

The Anisotropic Isolation Forest (AIF) enables anomaly detection with controllable feature sensitivities by introducing a novel isolation mechanism based on random hyperplanes that are more likely oriented along specified directions in the feature space. The AIF partitions the data using random hyperplanes whose normal vectors are sampled from anisotropic distributions. By appropriately modifying the anisotropy of these distributions, one can design AIF models that exhibit complex
sensitivity patterns across the feature space, with increased
sensitivity in some directions and reduced sensitivity in others.

The proposed method can be defined by Algorithms \ref{alg:aif_1}-\ref{alg:aif_3} inspired by \cite{Hariri, Liu}. There are no conceptual changes in Algorithms \ref{alg:aif_1} and \ref{alg:aif_3} (they are the same as in \cite{Hariri}), but we show them for the completeness of the exposition.

\begin{algorithm}[H]
\caption{Anisotropic iForest($X$, $t$, $c$, $\mathcal{D}$)}
\label{alg:aif_1}
\begin{algorithmic}[1]
\REQUIRE $X$ -- input data, $t$ -- number of trees, $c$ -- sub-sampling size, $\mathcal{D}$ -- distribution.
\ENSURE a set of $t$ Anisotropic iTrees
\STATE {\bf Initialize} \textit{Anisotropic Forest} $\leftarrow \emptyset$
\STATE Set height limit $l \leftarrow \lceil \log_2 c \rceil$
\FOR{$i = 1$ to $t$}
    \STATE $X' \leftarrow \text{sample}(X, c)$
    \STATE \textit{Anis. Forest} $\leftarrow$ \textit{Anis. Forest} $\cup$ $\text{Anis. iTree}(X', 0, l, \mathcal{D})$
\ENDFOR
\end{algorithmic}
\end{algorithm}

\begin{algorithm}[H]
\caption{ Anisotropic iTree$(X,e,l,\mathcal{D})$}
\label{alg:aif_2}
\begin{algorithmic}[1]
\REQUIRE $X$ -- input data, $e$ -- current tree height, $l$ -- height limit, $\mathcal{D}$ -- distribution ($N(\bm{0},\bm{A})$ or $\sum_{i=1}^p \pi_i N(\bm{0},\bm{A}_i)$).
\ENSURE an Anisotropic iTree
\IF{$e \ge l$ \OR $|X| \le 1$}
    \RETURN \textit{exNode}\{ \textit{Size: } $\leftarrow$
 $|X|$ \}
\ELSE
    \STATE Sample a normal vector $\mathbf{\bm{\omega}} \in \mathbb{R}^{\dim\{X\}}$ ($\dim\{X\}$ is a number of features in $X$) from the distribution~$\mathcal{D}.$
    \STATE Randomly select an intercept point $\mathbf{p} \in \mathbb{R}^{\dim\{X\}}$ within the range of $X$.
    \STATE Set components of $\mathbf{\bm{\omega}}$ to zero according to the chosen extension level.
    \STATE $X_l \leftarrow \text{filter}\big(X, (X - \mathbf{p})' \mathbf{\bm{\omega}} \le 0\big)$
    \STATE $X_r \leftarrow \text{filter}\big(X, (X - \mathbf{p})' \mathbf{\bm{\omega}} > 0\big)$
    \RETURN \\ \textit{inNode}\{Left: $\text{Anisotropic iTree}(X_l, e+1, l,\mathcal{D})$; \\
        \hspace{10.5mm} Right: $\text{Anisotropic iTree}(X_r, e+1, l, \mathcal{D})$; \\
        \hspace{10.5mm} Normal: $\mathbf{\bm{\omega}}$; \\
        \hspace{10.5mm} Intercept: $\mathbf{p}$
    \}
\ENDIF
\end{algorithmic}
\end{algorithm}

\begin{algorithm}[H]
\caption{PathLength($\mathbf{x}$, $T$, $e$)}
\label{alg:aif_3}
\begin{algorithmic}[1]
\REQUIRE $\mathbf{x}$ -- an instance, $T$ -- an Anisotropic iTree, $e$ -- current path length (initialized to $0$ when first called)
\ENSURE Path length of $\mathbf{x}$
\IF{$T$ is an external node}
    \RETURN $e$  
\ENDIF
\STATE $\mathbf{\bm{\omega}} \leftarrow T.\text{Normal}$
\STATE $\mathbf{p} \leftarrow T.\text{Intercept}$
\IF{$(\mathbf{x} - \mathbf{p})'\mathbf{\bm{\omega}} \le 0$}
    \RETURN $\text{PathLength}(\mathbf{x}, T.\text{left}, e + 1)$
\ELSIF{$(\mathbf{x} - \mathbf{p})' \mathbf{\bm{\omega}} > 0$}
    \RETURN $\text{PathLength}(\mathbf{x}, T.\text{right}, e + 1)$
\ENDIF
\end{algorithmic}
\end{algorithm}

\section{Results and discussions}
\label{sec:num}

\subsection{Anomaly score maps}

\begin{figure}[htb!]
\centering
\subfloat[EIF]{\includegraphics[height = 1.in,width=1.in]{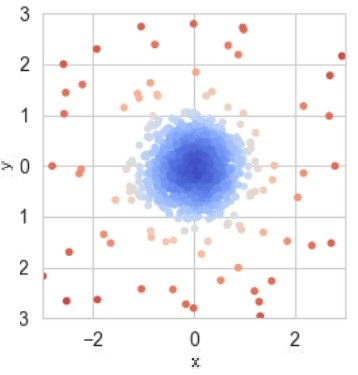}%
 \label{fig4.1}}
\hfil
\subfloat[AIF($\bm{A}_1$)]{\includegraphics[height = 1.in, width=1.in]{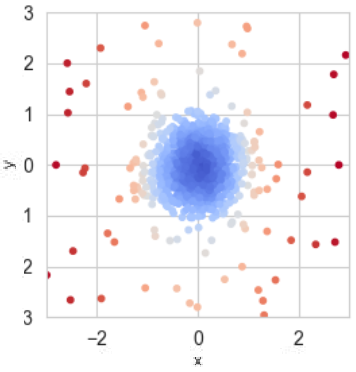}%
\label{fig4.2}}
\hfil
\subfloat[AIF($\bm{A}_2$)]{\includegraphics[ height = 1.in,width=1.2in]{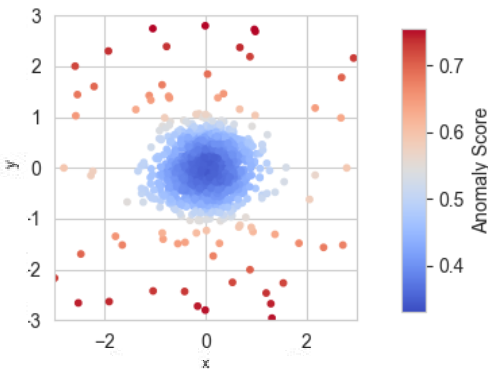}%
\label{fig4.3}}
\caption{Anomaly scores.}
\label{fig4}
\end{figure}

\begin{figure}[htb!]
\centering
\subfloat[EIF]{\includegraphics[height = 1.in, width=1.in]{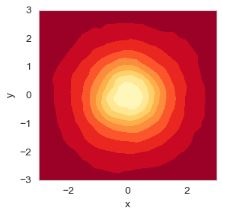}%
\label{fig5.1}}
\hfil
\subfloat[AIF($\bm{A}_1$)]{\includegraphics[height = 1.in, width=1.in]{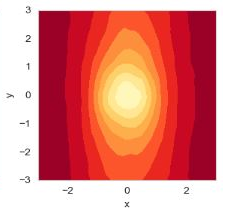}%
\label{fig5.2}}
\hfil
\subfloat[AIF($\bm{A}_2$)]{\includegraphics[height = 1.in,width=1.2in]{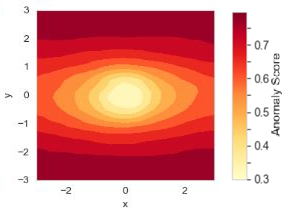}%
\label{fig5.3}}
\caption{Anomaly score maps.}
\label{fig5}
\end{figure}

As anomaly score maps are easier to visualize in two-dimensional space, we consider a 2D blob and assign to each of its points the anomaly scores returned by the EIF, AIF($\bm{A}_1$), and AIF($\bm{A}_2$), where $\bm{A}_1 = \mathrm{diag}(1, 0.05)$ and $\bm{A}_2 = \mathrm{diag}(0.05, 1)$. Each model consists of $500$ trees, with each tree built using a subsample of $128$ randomly selected data points. The data points, together with their anomaly scores from the three models, are shown in Fig.~\ref{fig4}, and the corresponding anomaly score maps are displayed in Fig.~\ref{fig5}.

For the EIF, the anomaly score map is isotropic: anomaly scores depend only on the distance from the center of the blob and are independent of direction in the feature space. The EIF treats deviations in the $x$- and $y$-coordinates equally, so data points with extreme $x$-values or $y$-values receive similar anomaly scores.

In contrast, AIF($\bm{A}_1$) is less sensitive to deviations in the $y$-coordinate. Data points with extreme $y$-values but nominal $x$-values receive lower anomaly scores than those assigned by the EIF. Note that AIF($\bm{A}_1$) still increases anomaly scores for points with extreme $y$-values, but to a lesser extent than for points with extreme $x$-values. Thus, AIF($\bm{A}_1$) does not ignore deviations in the $y$-coordinate; it is simply less sensitive to them. The contour lines of the anomaly scores produced by AIF($\bm{A}_1$) are elliptical, with axes proportional to the square roots of the entries of the matrix $\bm{A}_1$, and ellipses are therefore elongated along the $y$-axis. Analogously, AIF($\bm{A}_2$) is less sensitive to extreme $x$-values, and its anomaly score contours form ellipses elongated along the $x$-axis.

To further analyze how the above models assign anomaly scores in different directions of the feature space, for each model, we compute the average anomaly scores for data points lying along various directions. Define a ray in polar coordinates
\[
M(\theta) := \{(\rho,\theta') : \rho \ge 1,\ \theta' = \theta \}.
\]

\begin{figure}[htb!]
\centering
\subfloat[EIF]{\includegraphics[height = 1in, width=1.1in]{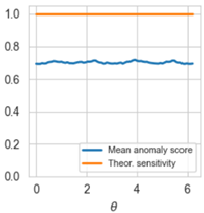}%
\label{fig6.1}}
\hfil
\subfloat[AIF($\bm{A}_1$)]{\includegraphics[height = 1.in, width=1.1in]{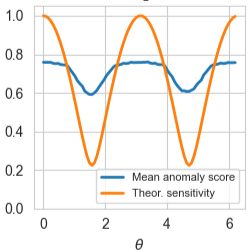}%
\label{fig6.2}}
\hfil
\subfloat[AIF($\bm{A}_2$)]{\includegraphics[height = 1.in, width=1.1in]{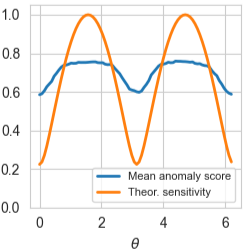}%
\label{fig6.3}}
\caption{Theoretical sensitivity $\alpha(\cdot)$ and mean anomaly scores.}
\label{fig6}
\end{figure}

For each anomaly score map in Fig.~\ref{fig5} and for each $\theta \in [0, 2\pi]$, we compute the mean anomaly score of the points in $M(\theta)$. Fig.~\ref{fig6} visualizes these mean anomaly scores together with the values of the measure of directional sensitivity \eqref{eq:sens1}, which is given in polar coordinates as
\begin{equation}
\label{sens_polar}
\alpha(\theta) = \sqrt{\bm{n}' \bm{A}\,\bm{n}}, \
\bm{n} = (\cos\theta,\sin\theta),\ \theta\in[0,2\pi],
\end{equation}
where the matrix $\bm{A}$ is chosen as $\bm{I}$, $\bm{A}_1$, and $\bm{A}_2$ for the EIF, AIF($\bm{A}_1$), and AIF($\bm{A}_2$), respectively.

For the EIF, the mean anomaly scores are identical across all directions in the feature space, and the measure of directional sensitivity is constant $\alpha(\theta)=1$ for all $\theta \in [0,2\pi]$. For the two AIF models, the behavior of $\alpha(\cdot)$ closely follows that of the mean anomaly scores. For AIF($\bm{A}_1$), $\alpha(\cdot)$ attains its smallest values at $\theta=\pi/2$ and $\theta=3\pi/2$, corresponding to directions aligned with the $y$-axis, and the mean anomaly scores are likewise lowest in these directions. Similarly, for AIF($\bm{A}_2$), $\alpha(\cdot)$ reaches its minimum at $\theta=0$ and $\theta=\pi$, corresponding to directions aligned with the $x$-axis, where the mean anomaly scores are also lowest.

\begin{figure}
\centering
\subfloat[EIF]{\includegraphics[height = 1.in,width=1.1in]{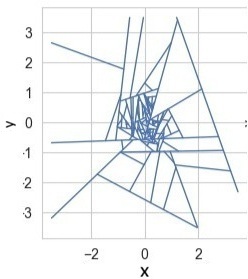}%
\label{fig7.1}}
\hfil
\subfloat[AIF($\bm{A}_1$)]{\includegraphics[height = 1.in,width=1.1in]{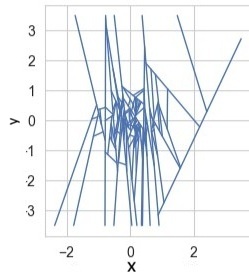}%
\label{fig7.2}}
\hfil
\subfloat[AIF($\bm{A}_2$)]{\includegraphics[height = 1.in,width=1.1in]{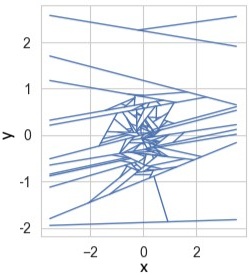}%
\label{fig7.3}}
\caption{Partitions of the feature space.}
\label{fig7}
\end{figure}

Fig.~\ref{fig7} shows the partitions of the feature space produced by the first search trees in the EIF, AIF($\bm{A}_1$), and AIF($\bm{A}_2$), respectively. Unlike the EIF, where the partition lines are isotropically oriented (the angles between the lines and the $x$-axis are uniformly distributed over $[0,\pi]$), the partition lines in AIF($\bm{A}_1$) are more closely aligned with the $y$-axis, while those in AIF($\bm{A}_2$) are more closely aligned with the $x$-axis. As a result, the AIF partitions exhibit directional bias: cells in AIF($\bm{A}_1$) are more elongated along the $y$-axis, whereas those in AIF($\bm{A}_2$) are more elongated along the $x$-axis. Thus, points along the $y$-axis are isolated more slowly in AIF($\bm{A}_1)$ partitions, while points along the $x$-axis are isolated more slowly in AIF($\bm{A}_2)$ partitions, resulting in lower anomaly scores for those points.

Now, let us consider AIF models constructed by the means of mixtures of Gaussian distributions. Let us put

\begin{equation*}
\bm{A}^{(1)}_1 = 
\begin{pmatrix}
1 & 0\\
0 & 0.01
\end{pmatrix},
\bm{A}^{(1)}_2 =
\begin{pmatrix}
0.01 & 0\\
0 & 1
\end{pmatrix} , 
\end{equation*}
\begin{equation*}
\bm{A}^{(2)}_1 = 
\begin{pmatrix}
1 & 0.99\\
0.99 & 1
\end{pmatrix}, 
\bm{A}^{(2)}_2 =
\begin{pmatrix}
1 & -0.99\\
-0.99 & 1
\end{pmatrix},
\end{equation*} and denote by AIF$(\mathcal{D}_1)$ the AIF model constructed using the mixture of Gaussian distributions
\[
\frac{1}{2}\bigl(N(\mathbf{0}, 2\mathbf{A}^{(1)}_1) + N(\mathbf{0}, 2\mathbf{A}^{(1)}_2)\bigr),
\]
and by AIF$(\mathcal{D}_2)$ the AIF model constructed using the distribution
\[
\frac{1}{2}\bigl(N(\mathbf{0}, 2\mathbf{A}^{(2)}_1) + N(\mathbf{0}, 2\mathbf{A}^{(2)}_2)\bigr).
\]
In the above mixtures, we multiply the covariance matrices by $2$ so that in both cases the normalisation condition \eqref{eq:sens_adj} is satisfied. This normalisation does not change the performance of the constructed models.

\begin{figure}[htb!]
\centering
\subfloat[AIF$(\mathcal{D}_1)$]{\includegraphics[height = 1.2in, width=1.5in]{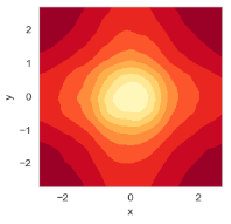}%
\label{fig8.1}}
\hfil
\subfloat[AIF$(\mathcal{D}_2)$]{\includegraphics[height = 1.2in, width=1.8in]{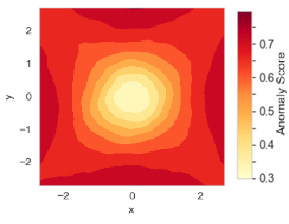}%
\label{fig8.2}}
\caption{Anomaly score maps.}
\label{fig8}
\end{figure}

Fig.~\ref{fig8} shows the anomaly score maps produced by AIF$(\mathcal{D}_1)$ and AIF$(\mathcal{D}_2)$. For AIF$(\mathcal{D}_1)$, points with extreme values in either the $x$- or $y$-coordinate are assigned relatively lower anomaly scores, while points exhibiting simultaneous extremes in both coordinates receive higher anomaly scores. This indicates that AIF$(\mathcal{D}_1)$ is less sensitive to deviations in a single coordinate but more sensitive to those in both coordinates. The opposite behavior is observed for AIF$(\mathcal{D}_2)$.

\begin{figure}[htb!]
\centering
\hspace{3mm}
\subfloat[AIF$(\mathcal{D}_1)$]{\includegraphics[height = 1.in,width=1.3in]{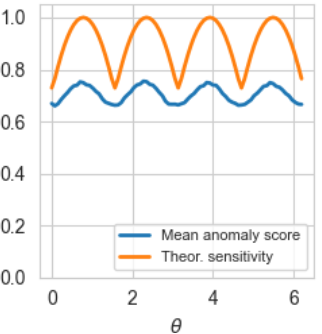}%
\label{fig9.1}}
\hspace{3mm}
\subfloat[AIF$(\mathcal{D}_2)$]{\includegraphics[height = 1.in,width=1.3in]{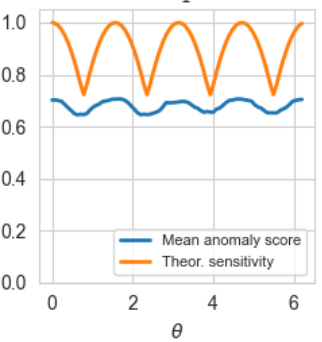}%
\hspace{7mm}
\label{fig9.2}}
\caption{Theoretical sensitivities $\widetilde\alpha(\cdot)$ and mean anomaly scores.}
\label{fig9}
\end{figure}

Fig.~\ref{fig9} shows the mean anomaly scores in the anomaly score maps from Fig.~\ref{fig8}, computed for the points in $M(\theta),\ \theta \in [0, 2\pi]$, together with the values of measure of directional sensitivity $\widetilde\alpha(\theta),\ \theta \in [0, 2\pi]$ defined in \eqref{eq:sens11}. For AIF$(\mathcal{D}_1)$, it holds
\[
\widetilde\alpha(\theta) = \tfrac{\sqrt{2}}{2}\left(\sqrt{\bm{n}' \bm{A}^{(1)}_1 \bm{n}} + \sqrt{\bm{n}' \bm{A}^{(1)}_2 \bm{n}}\right),
\bm{n} = (\cos\theta,\sin\theta),
\]
and an analogous expression is used for AIF$(\mathcal{D}_2)$, with the matrices $\bm{A}^{(2)}_1$ and $\bm{A}^{(2)}_2$. One can see that, for both models, the mean anomaly scores are aligned with the behaviour of $\widetilde\alpha(\cdot),$ so that both models assign anomaly scores consistently with the prescribed sensitivity.

\begin{figure}[htb!]
\centering
\subfloat[AIF$(\mathcal{D}_1)$]{\includegraphics[height = 1.2in, width=1.3in]{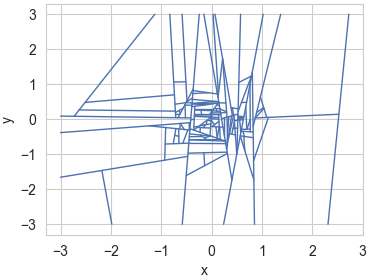}%
\label{fig10.1}}
\hspace{3mm}
\subfloat[AIF$(\mathcal{D}_2)$]{\includegraphics[height = 1.2in,width=1.3in]{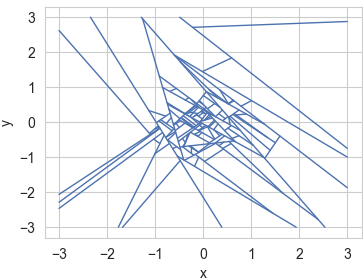}%
\hspace{7mm}
\label{fig10.2}}
\caption{Partitions of the feature space.}
\label{fig10}
\end{figure}

Fig.~\ref{fig10} shows the partitions produced by the first search trees in AIF$(\mathcal{D}_1)$ and AIF$(\mathcal{D}_2)$. In the case of AIF$(\mathcal{D}_1)$, the partition lines are more closely aligned with the coordinate axes, whereas in AIF$(\mathcal{D}_2)$ they are more frequently oriented along the diagonals of $\mathbb{R}^2$. As a result, the cells in the two partitions are elongated along the axes and along the diagonals, respectively. Thus, points located along the axes are isolated more slowly in AIF$(\mathcal{D}_1)$ partitions, while points located along the diagonals are isolated slowlier in AIF$(\mathcal{D}_2)$ partitions, leading to lower anomaly scores assigned to those points.


\subsection{Multidimensional data}

Let us study the performance of the AIF in a multidimensional setting.  We construct several AIF models under various feature sensitivity scenarios and compare their performance with that of the standard EIF.

First, for the Diabetes dataset \cite{diabetes} consider a scenario in which selected features are assigned greater sensitivity. Suppose one wishes to construct a model that is more sensitive to the features \textit{age} and \textit{diabetes pedigree function} (\textit{DPF}, the likelihood of an individual developing diabetes), such that the model is more sensitive to a higher-risk group consisting of individuals with extreme values of these features. Consider AIF$(\bm{A}_3)$, where
\(
\bm{A}_3 = \tfrac{1}{1000}\,\mathrm{diag}(1000, 1000, 1, \ldots, 1),
\)
so that the features \textit{age} and \textit{DPF} are assigned greater sensitivity, while all other features retain standard sensitivity. The multiplier $\tfrac{1}{1000}$ does not affect the model’s performance but is included to normalize the spectral radius of $\bm{A}_3$ to $1$. 

We construct the EIF and AIF$(\bm{A}_3)$ models, each consisting of $500$ search trees, with each tree trained on a subsample of $128$ randomly selected data points. For each model, we mark as anomalies the data points whose anomaly scores fall within the top $10\%$ of all scores produced by that model.

\begin{figure}[htb!]
\centering
\subfloat{\includegraphics[height = 1.in, width=1.5in]{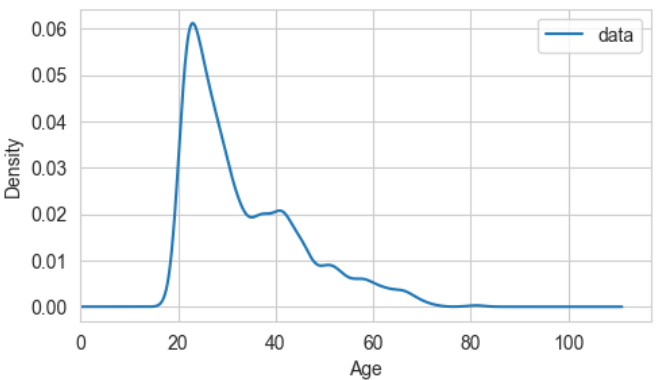}%
\label{fig_age}}
\hfil
\subfloat{\includegraphics[height = 1.in, width=1.5in]{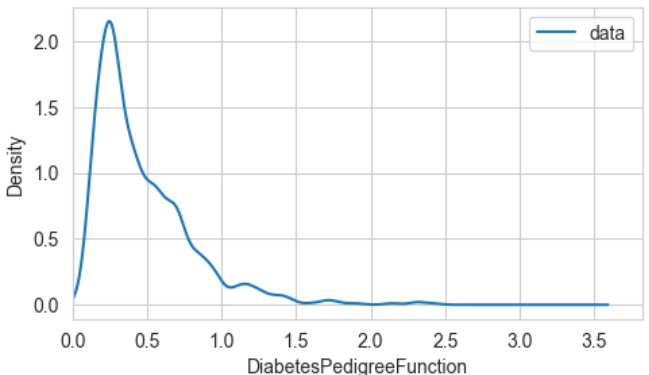}%
\label{fig_dpf}}
\caption{Data distribution.}
\label{fig_data}
\end{figure}

\begin{figure}[htb!]
\centering
\subfloat{\includegraphics[height = 1.2in, width=1.5in]{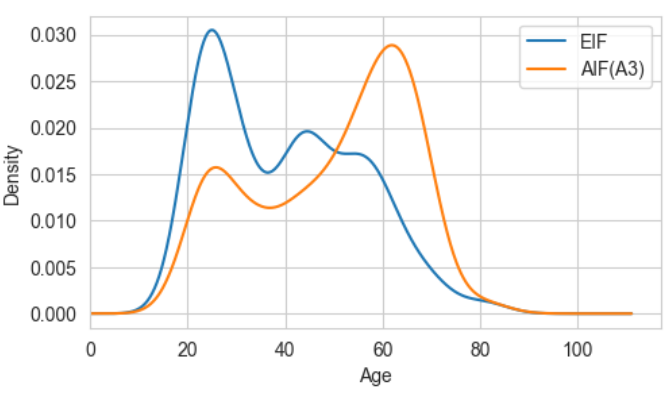}%
\label{fig11.1}}
\hfil
\subfloat{\includegraphics[height = 1.2in, width=1.5in]{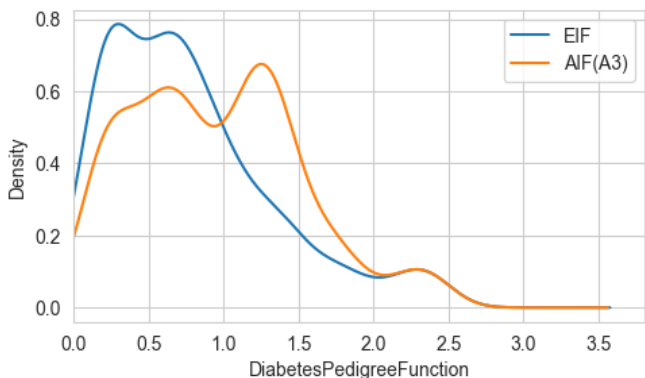}%
\label{fig11.2}}
\caption{Distributions of anomalies found by AIF$(\bm{A}_3)$.}
\label{fig11}
\end{figure}

Fig.~\ref{fig_data} shows the distribution of all observations across the features \textit{age} and \textit{DPF}, while Fig.~\ref{fig11} shows the distributions of the anomalies identified by the EIF and AIF$(\bm{A}_3)$ across these features. Compared with the EIF, AIF$(\bm{A}_3)$ identifies a larger number of anomalies with high values in these features, which is consistent with its increased sensitivity to them.

We compare the mean values of the features of the anomalies identified by the EIF and AIF$(\bm{A}_3)$ using a $t$-test for independent samples. The resulting $p$-values are reported in the first column of Table~\ref{table1}. For \textit{age} and \textit{DPF}, the $t$-test yields $p$-values below $0.05$, indicating that AIF$(\bm{A}_3)$ finds anomalies with significantly higher values of these features,confirming its increased sensitivity to them. 

\begin{figure}[htb!]
\centering
\includegraphics[height = 2in, width=3in]{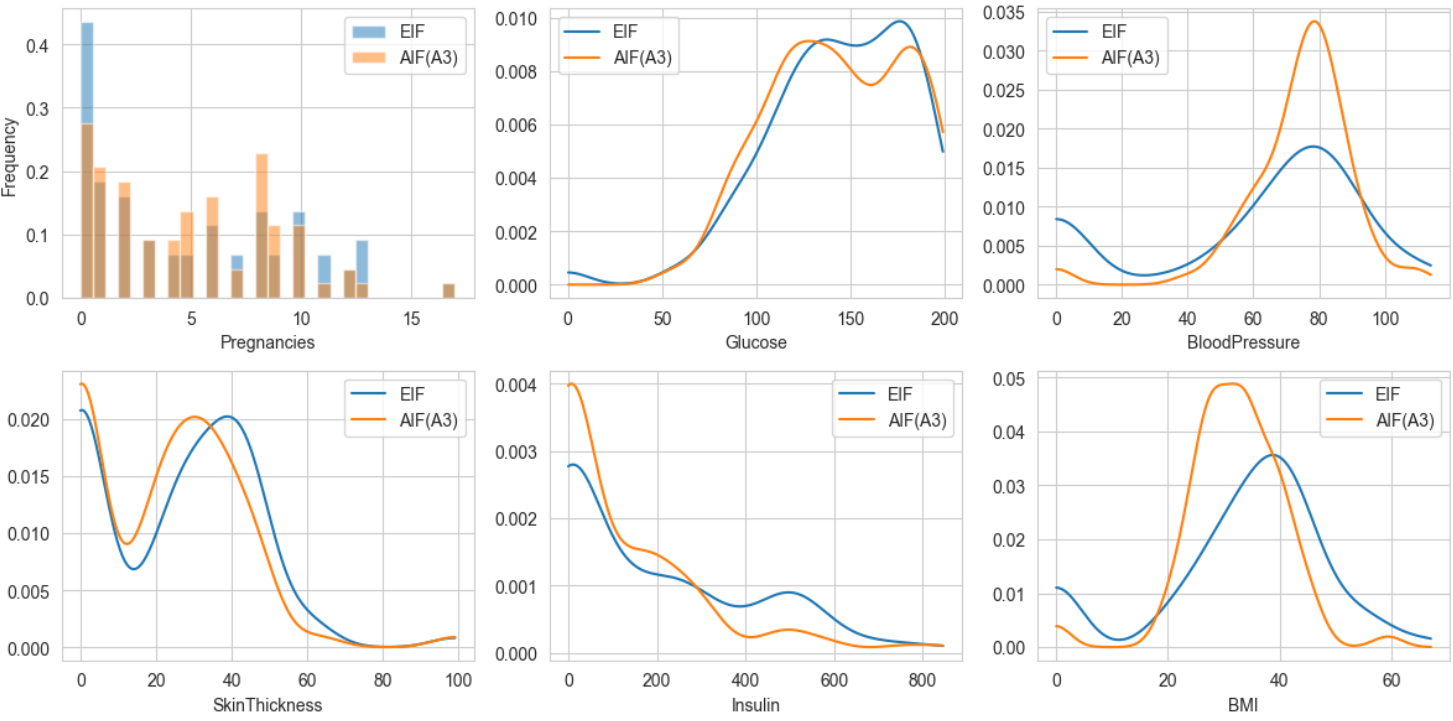}%
\caption{Distributions of anomalies across features with standard AIF$(\bm{A}_3)$ sensitivity.}
\label{fig12}
\end{figure}

Fig.~\ref{fig12} shows the distributions of anomalies for the remaining features, for which standard sensitivity was assigned in AIF$(\bm{A}_3)$. The anomaly distributions produced by EIF and AIF$(\bm{A}_3)$ also differ for \textit{blood pressure}, which occurs due to correlations between this feature and \textit{age} and \textit{DPF}. For the remaining features, which exhibit weaker correlations with \textit{age} and \textit{DPF}, no statistically significant differences are observed in the mean values of the anomalies identified by the EIF or AIF$(\bm{A}_3)$.

\begin{table}[h!]
\caption{$T$-test $p$-values.}
\begin{tabular}{lccc}
\toprule
\textbf{Feature} & \textbf{AIF$(\bm{A}_3)$} & \textbf{AIF$(\bm{A}_4)$} & \textbf{AIF$(\bm{A}_5)$} \\
\midrule
Age & 0.01 & 0.01 & 0.05 \\
DiabetesPedigreeFunction & 0.05 & 0.01 & 0.01 \\
Pregnancies & 0.90 & 0.59 & 0.53 \\
Glucose & 0.91 & 0.16 & 0.33 \\
BloodPressure & 0.01 & 0.01 & 0.95 \\
SkinThickness & 0.34 & 0.10 & 0.66 \\
Insulin & 0.14 & 0.52 & 0.57 \\
BMI & 0.61 & 0.49 & 0.98 \\
\bottomrule
\end{tabular}
\label{table1}
\end{table}

Let us now consider the reduced-sensitivity scenario. Consider AIF$(\bm{A}_4)$, where 
\(
\bm{A}_4 = \mathrm{diag}(0.001, 0.001, 1, \ldots, 1),
\)
so that the features \textit{age} and \textit{DPF} are assigned lower sensitivity, while all other features retain standard sensitivity.

\begin{figure}[htb!]
\centering
\subfloat{\includegraphics[height = 1.2in, width=1.65in]{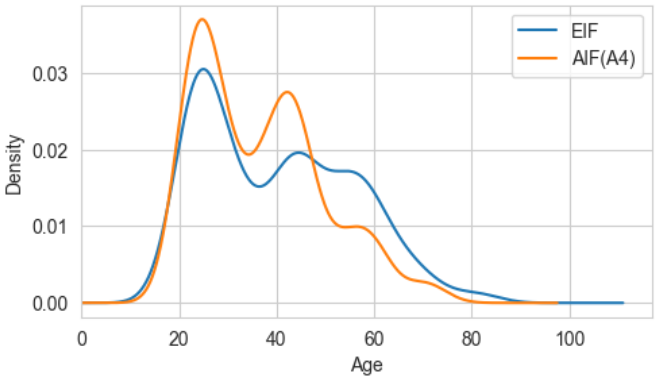}%
\label{fig13.1}}
\hfil
\subfloat{\includegraphics[height = 1.2in, width=1.65in]{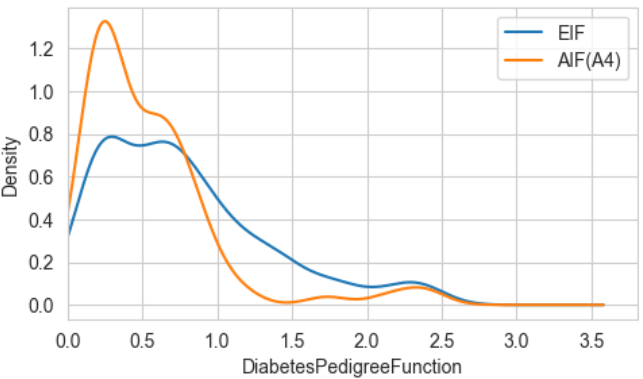}%
\label{fig13.2}}
\caption{Distributions of anomalies found by AIF$(\bm{A}_4)$.}
\label{fig13}
\end{figure}

Fig.~\ref{fig13} shows the distributions of anomalies identified by the EIF and AIF$(\bm{A}_4)$ across the features \textit{age} and \textit{DPF}. Compared to EIF, AIF$(\bm{A}_4)$ identifies fewer anomalies with high values of these features, which is consistent with its reduced sensitivity to them. The second column of Table~\ref{table1} reports the $p$-values of the $t$-tests that compare the mean values of the features of the anomalies detected by the EIF and AIF$(\bm{A}_4)$. As expected, AIF$(\bm{A}_4)$ identifies anomalies with statistically significantly lower \textit{age} and~\textit{DPF}.

\begin{figure}[htb!]
\centering
\subfloat{\includegraphics[height = 1.2in, width=1.5in]{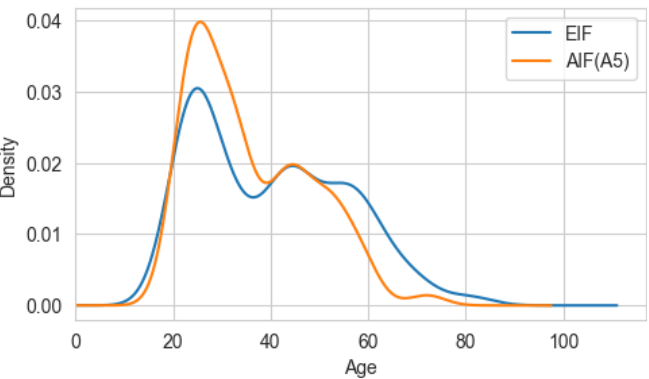}%
\label{fig14.1}}
\hfil
\subfloat{\includegraphics[height = 1.2in, width=1.5in]{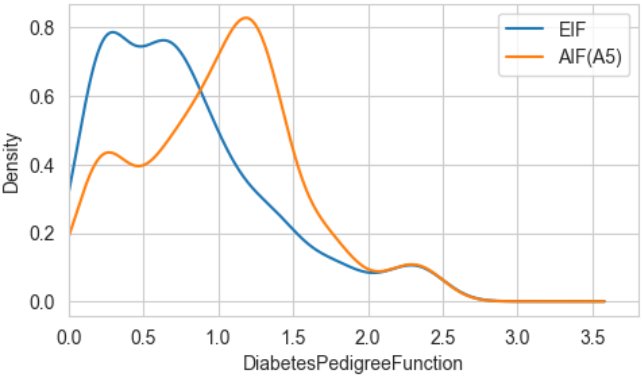}%
\label{fig14.2}}
\caption{Distributions of anomalies found by AIF$(\bm{A}_5)$.}
\label{fig14}
\end{figure}

We now consider a mixed-sensitivity scenario. In AIF$(\bm{A}_5)$, where 
\(
\bm{A}_5 = \tfrac{1}{100}\mathrm{diag}(0.01, 100, 1, \ldots, 1),
\)
the feature \textit{age} is assigned a lower sensitivity, while the feature \textit{DPF} is assigned a higher sensitivity; all other features retain standard sensitivity. Fig.~\ref{fig14} shows the distributions of anomalies identified by the EIF and AIF$(\bm{A}_5)$ across the features \textit{age} and \textit{DPF}. Compared to EIF, AIF$(\bm{A}_5)$ identifies fewer anomalies with large \textit{age} values and more anomalies with large \textit{DPF} values. The  third column of Table~\ref{table1} reports the $p$-values of the $t$-tests that compare the mean feature values of the anomalies detected by the EIF and AIF$(\bm{A}_5)$. These results confirm that AIF$(\bm{A}_5)$ modifies  sensitivity to \textit{age} and \textit{DPF} as intended.

A Python implementation of the algorithm is available at \cite{link}, along with a Jupyter notebook that reproduces the figures and experiments presented in this paper.

\section{Acknowledgement}
This study was supported by Ripple Impact Fund 2022-247584 (5855) and by the Australian Research Council's Discovery Projects funding scheme (project DP220101680).

\appendix
\label{sect:append}
Let $\bm{A}$ be a symmetric positive definite $d\times d$ matrix and the ellipsoid $E$ is defined as 
\[\bm{x}'\bm{A}^{-1}\bm{x}=1.\] Let $\bm{n}\in\mathbb{S}^{d-1},$ then the half-length of the projection of $E$ onto the axis directed by $\bm{n}$ satisfies $\sqrt{\bm{n}' \bm{A}^{-1} \bm{n}}.$

Indeed, the length of a projection of a vector $\bm{x}$ onto the axis directed by a unit length vector $\bm{n}$ satisfies $||\bm{x}||\cos (\bm{x},\bm{n}) = \bm{x}'\bm{n}.$ Thus, one needs to maximize $\bm{x}'\bm{n}$ over the ellipsoid $\bm{x}'\bm{A}^{-1}\bm{x}=1.$ Form the Lagrangian

\[ \mathcal{L}(\bm{x},\lambda) = \bm{x}'\bm{n} - \lambda(\bm{x}'\bm{A}^{-1}\bm{x}-1).\] Then, it holds
\[  \frac{\partial \mathcal{L}(\bm{x},\lambda)}{\partial \bm{x}} = \bm{n} -2\lambda \bm{A}^{-1}\bm{x} = 0, \ {\rm when} \ \bm{x} = \frac{1}{2\lambda}\bm{A}\bm{n}.\] By putting the above into the constrain $\bm{x}'\bm{A}^{-1}\bm{x}=1,$ one gets
\[ \left( \frac{1}{2\lambda} \right)^2 \bm{n}'\bm{A}\bm{n} =1, \ {\rm so\ that} \ \lambda = \pm\frac{1}{2}\sqrt{\bm{n}'\bm{A}\bm{n}}.\] Thus, the maximising point is 

\[ \bm{x}_{max} =  \frac{\bm{A}\bm{n}}{\sqrt{\bm{n}'\bm{A}\bm{n}}}.\] As $\bm{A}$ is symmetric, it holds $\bm{x}'_{max}\bm{n} = \sqrt{\bm{n}'\bm{A}\bm{n}}.$

\bibliographystyle{abbrv}
\bibliography{sample-base}

@String{Computing = "Computing" }

@String{Computer = "{IEEE} Computer" }

@String{Springer = "Springer-Verlag" }

@book{abr,
  author    = {Abramowitz, M. and Stegun, I.},
  title     = {Handbook of Mathematical Functions with Formulas, Graphs, and Mathematical Tables},
  publisher = {US Government Printing Office},
  address   = {Washington},
  year      = {1968}
}

@book{Barnett,
  author    = {Barnett, V. and Lewis, T.},
  title     = {Outliers in Statistical Data},
  publisher = {Wiley},
  address   = {New York},
  year      = {1994}
}

@inproceedings{Breunig,
  author    = {Breunig, M. and Kriegel, H. P. and Ng, R. T. and Sander, J.},
  title     = {LOF: Identifying Density-Based Local Outliers},
  booktitle = {Proceedings of the 2000 ACM SIGMOD International Conference on Management of Data},
  pages     = {93--104},
  year      = {2000}
}

@book{Eskin,
  author    = {Eskin, E. and Arnold, A. and Prerau, M. and Portnoy, L. and Stolfo, S.},
  title     = {Applications of Data Mining in Computer Security},
  publisher = {Springer},
  address   = {Boston},
  year      = {2002}
}

@article{Hariri,
  author    = {Hariri, S. and Kind, M. C. and Brunner, R. J.},
  title     = {Extended Isolation Forest},
  journal   = {IEEE Transactions on Knowledge and Data Engineering},
  volume    = {33},
  number    = {4},
  pages     = {1479--1489},
  year      = {2021}
}

@article{kmeans,
  author    = {Karczmarek, P. and Kiersztyn, A. and Pedrycz, W. and Al, E.},
  title     = {K-means-Based Isolation Forest},
  journal   = {Knowledge-Based Systems},
  volume    = {195},
  pages     = {105659},
  year      = {2020}
}

@inproceedings{Liu,
  author    = {Liu, F. T. and Ting, K. M. and Zhou, Z. H.},
  title     = {Isolation Forest},
  booktitle = {2008 Eighth IEEE International Conference on Data Mining},
  pages     = {413--422},
  year      = {2008}
}

@inproceedings{Rasmussen,
  author    = {Rasmussen, C. E.},
  title     = {The Infinite {G}aussian Mixture Model},
  booktitle = {Proceedings of the 12th International Conference on Neural Information Processing Systems},
  pages     = {554--560},
  year      = {1999}
}

@article{Xu,
  author    = {Xu, H. and Pang, G. and Wang, Y. and Wang, Y.},
  title     = {Deep Isolation Forest for Anomaly Detection},
  journal   = {IEEE Transactions on Knowledge and Data Engineering},
  volume    = {35},
  number    = {12},
  pages     = {12591--12604},
  year      = {2023}
}

@misc{diabetes,
  title     = {Diabetes Dataset},
  howpublished = {\url{https://www.kaggle.com/datasets/mathchi/diabetes-data-set}}
}

@misc{link,
  title     = {GitHub link},
  howpublished = {\url{https://github.com/elliot1776/AIF/tree/main}}
}

@ArtifactSoftware{R,
    title = {R: A Language and Environment for Statistical Computing},
    author = {{R Core Team}},
    organization = {R Foundation for Statistical Computing},
    address = {Vienna, Austria},
    year = {2019},
    url = {https://www.R-project.org/},
}

\end{document}